\documentclass[pra,showpacs,preprintnumbers,amsmath,amssymb,superscriptaddress,nofootinbib]{revtex4-1}

\usepackage{amsmath,amssymb,times,bm}
\usepackage[colorlinks,linkcolor=red]{hyperref}

\newcommand{\be}{\begin{equation}}
\newcommand{\ee}{\end{equation}}
\newcommand{\ben}{\begin{eqnarray}}
\newcommand{\een}{\end{eqnarray}}
\newcommand{\bF}{\begin{figure}}
\newcommand{\eF}{\end{figure}}

\newcommand{\bes}{\begin{subequations}}
\newcommand{\ees}{\end{subequations}}
\newtheorem{theorem}{Theorem}

\long\def\symbolfootnote[#1]#2{\begingroup%
\def\thefootnote{\fnsymbol{footnote}}\footnote[#1]{#2}\endgroup}

\def\ket#1{ | #1 \rangle}
\def\bra#1{{\langle #1 | }}
\def\tr{ {\rm{Tr }}}

\newcommand{\proj}[1]{\mbox{$|#1\rangle \!\langle #1 |$}}

\begin{document}
\title{Role of quantum discord in quantum communication\symbolfootnote[1]{To be published in Springer's Lecture Notes in Computer Science as proceedings of The sixth Conference on the Theory of Quantum Computation, Communication and Cryptography (TQC,2011) under the title \textit{Quantum discord in quantum information theory - From strong subadditivity to the Mother protocol}}}

%\title{Quantum discord in quantum information theory - From strong subadditivity to the Mother protocol}
%
\author{Vaibhav Madhok}
\affiliation{Center for Quantum Information and Control, University of New Mexico, Albuquerque, NM 87131-0001, USA}

\author{Animesh Datta}
\affiliation{Clarendon Laboratory, Department of Physics, University of Oxford, OX1 3PU, United Kingdom}

\begin{abstract}
Positivity of quantum discord is shown to be equivalent to the strong sub additivity of the Von-Nuemann entropy. This leads to a connection between the mother protocol of quantum information theory~[A. Abeyesinghe, I. Devetak, P. Hayden, and A. Winter, Proc. R. Soc. A \textbf{465}, 2537, (2009)] and quantum discord. We exploit this to show that quantum discord is a measure coherence in the performance of the mother protocol. Since the mother protocol is a unification of an important class of problems (those that are bipartite, unidirectional and memoryless), we show quantum discord to be a measure of coherence in these protocols. Our work generalizes an earlier operational interpretation of quantum discord provided in terms of quantum state merging~[V. Madhok and A. Datta, Phys. Rev. A, \textbf{83}, 032323, (2011)].
\end{abstract}

\maketitle

\section{\label{sec:Intro}Introduction}

Why quantum mechanics provides enhancements and speedups over best known classical procedures forms one of the most fundamental questions in quantum information science. This has canonically been answered in terms of quantum entanglement~\cite{pv07}. This is however far from complete.There are quantum processes which provide an exponential advantage in the presence of little or no entanglement~\cite{dqc1}. In the realm of mixed-state quantum computation, quantum discord~\cite{z00,oz02} has been proposed as resource~\cite{dsc08} and there are already some formal proofs in that direction~\cite{e10}. The role of quantum discord in quantum information theory, however, still remains unclear. Recently, operational interpretations for quantum discord have been provided in terms of the quantum state merging protocol~\cite{md10,cabmpw10}. In this paper, we go beyond this by exhibiting the role of quantum discord in essentially \emph{all} quantum information processing protocols.

Quantum discord aims at capturing all quantum correlations in a quantum state, including entanglement~\cite{z00,oz02,hv01}. Quantum mutual information is generally taken to be the measure of total correlations, classical and quantum, in a quantum state. For two systems, $A$ and $B$, it is defined as $ I(A:B) = S(A) + S(B) -S(A,B),$ where $S(\cdot)$ stands for the von Neumann entropy, $S(\rho)\equiv -\tr(\rho\log\rho).$ In our paper, all logarithms are taken to base 2. For a classical probability distribution, Bayes' rule leads to an equivalent definition of the mutual information as $I(A:B) = S(A)-S(A|B),$ where the conditional entropy $S(A|B)$ is an average of the Shannon entropies of $A,$ conditioned on the alternatives of $B.$ It captures the ignorance in $A$ once the state of $B$ has been determined. For a quantum system, this depends on the measurements that are made on $B.$ For a POVM given by the set $\{\Pi_i\},$ the state of $A$ after the measurement corresponding to the outcome $i$ is given by
\be
\label{eq:postmeas}
\rho_{A|i} = \tr_B(\Pi_i\rho_{AB})/p_i,\;\;\;p_i=\tr_{A,B}(\Pi_i\rho_{AB}).
\ee
A quantum analogue of the conditional entropy can then be defined as $\tilde{S}_{\{\Pi_i\}}(A|B) \equiv \sum_i p_i S(\rho_{A|i}),$ and an alternative version of the quantum mutual information can now be defined as $\mathcal{J}_{\{\Pi_i\}}(A:B) = S(A)-\tilde{S}_{\{\Pi_i\}}(A|B).$ The above quantity depends on the chosen set of measurements $\{\Pi_i\}.$ To capture all the classical correlations present in $\rho_{AB},$ we maximize $\mathcal{J}_{\{\Pi_i\}}(A:B)$ over all $\{\Pi_i\},$ arriving at a measurement independent quantity $\mathcal{J}(A:B) = \max_{\{\Pi_i\}}(S(A)-\tilde{S}_{\{\Pi_i\}}(A|B)) \equiv S(A)-\tilde{S}(A|B),$ where $\tilde{S}(A|B)=\min_{\{\Pi_i\}}\tilde{S}_{\{\Pi_i\}}(A|B).$
Since the conditional entropy is concave over the set of POVMs, which is convex, the minimum is attained on the extreme points of the set of POVMs, which are rank 1~\cite{dattathesis}. Then, quantum discord is finally defined as
\ben
\label{discexp}
\mathcal{D}(A:B) &=& I(A:B)-\mathcal{J}(A:B) \\
                 &=& S(A)-S(A:B)+\min_{\{\Pi_i\}}\tilde{S}_{\{\Pi_i\}}(A|B),\nonumber
\een
where $\{\Pi_i\}$ are now, and henceforth in the paper, rank 1 POVMs. It is well known that the quantum discord is non-negative for all quantum states~\cite{oz02,dattathesis}.

Our endeavour in this paper shall be to clarify the role of quantum discord in quantum information theory. We shall show how quantum discord quantifies the coherence in a large class of quantum resource inequalities, beginning with the so-called `mother protocol'. Our way into the heart of quantum information theory will be though strong sub-additivity (SSA) of von-Neumann entropy. SSA is one of the most fundamental inequalities in information theory, quantum and classical. We shall use a simple proof of SSA provided by quantum state merging protocol~\cite{how07}, to lead us onto the mother protocol of quantum information theory~\cite{adhw09}. The mother protocol which can be thought of as the fully quantum Slepian-Wolf protocol, achieving quantum communication-assisted entanglement distillation, and state transfer from the sender to the receiver. It also has as its children several important and common protocols like quantum teleportation and entanglement distillation. We shall discuss the role of quantum discord in the performance of the noisy versions these protocols in the final section of our paper.

\section{Quantum discord and strong subadditivity}
We begin by providing a new proof of the positivity of quantum discord~\cite{oz02} by casting it in terms of the SSA of von-Neumann entropy.
\begin{theorem}
 \label{discordispositove} Strong subadditivity of the Von Neumann entropy implies nonnegativity of the quantum discord.
\end{theorem}

\noindent {\it Proof: } Consider the joint state $\rho_{AB}$ subject to one dimensional orthogonal measurements $\Pi_j=\proj{e_j}$ on $B$, extended to arbitrary (at most $dim(B)^2$) dimensions. Then
 $$
 p_j\,\rho_{A|j} = \tr_{B}(\rho_{AB}\Pi_j)=\bra{e_j}\rho_{AB}\ket{e_j},\;\;\;  p_j = \tr_{B}(\rho_{B}\Pi_j)=\bra{e_j}\rho_{B}\ket{e_j}.
 $$
Note that the measurement is made on the system $B$, while in the definition of quantum discord, it was on $A$. Quantum discord is not symmetric under the exchange of the subsystems, but this is not a concern as we just as well have proved the result for $\mathcal{D}(B,A)$.

Suppose now that a system $C$ interacts with $B$ so as to make the desired measurement ($U\ket{e_j}\otimes\ket{0}=\ket{e_j}\otimes\ket{f_j}$), leaving the state
\be
\label{extrho}
\rho'_{ABC}=\sum_{j,k}\bra{e_j}\rho_{AB}\ket{e_k}\otimes\ket{e_j}\bra{e_k}\otimes\ket{f_j}\bra{f_k}.
\ee
If the eigendecomposition of $ \rho_{AB} =\sum_l \lambda_l\proj{r_l} $, then
 \be
\rho'_{ABC}=\sum_{j,k,l}\lambda_l\langle \mathbb{I}_A,e_j\proj{r_l}\mathbb{I}_A,e_k\rangle\otimes\ket{e_j}\bra{e_k}\otimes\ket{f_j}\bra{f_k}
= \sum_{l}\lambda_l\proj{e_l,r_l,f_l}\nonumber
\ee whereby
 $$
 S(\rho'_{ABC})=S(\rho_{AB}).
 $$
Also, from Eq. (\ref{extrho}),
 \bes
\label{partials}
 \ben
 \rho'_{AB}&=&\sum_{j}p_j\rho_{A|j}\otimes\proj{e_j},\;\;\;\mbox{so}\;\;\;S(\rho'_{AB})=S({\bm p})+\sum_j p_j S(A|j),\\
 \rho'_{BC}&=&\sum_{j,k}\proj{e_j}\rho_B\proj{e_k}\otimes\ket{f_j}\bra{f_k},\;\;\;\mbox{so}\;\;\;S(\rho'_{BC})=S(\rho_B),\\
  \rho'_{B}&=&\sum_{j}p_j\proj{e_j},\;\;\;\mbox{so}\;\;\;S(\rho'_{B})=S({\bm p}).
 \een
 \ees
Now use the strong subadditivity of the von-Neumann entropy~\cite{lr73} which is
 \be
\label{ssa}
S(\rho'_{ABC}) + S(\rho'_{B}) \leq S(\rho'_{AB}) + S(\rho'_{BC}).
\ee
Eqs. (\ref{partials}) reduce this to
\be
S(\rho_{AB}) + S({\bm p}) \leq S({\bm p})+\sum_j p_j S(A|j) + S(\rho_B),
\ee
whereby
\be
\tilde S_{\{\Pi_j\}}(A|B) \equiv \sum_j p_j S(A|j)\geq S(\rho_{AB})- S(\rho_B) \equiv S(A|B).
\ee
This, being true for all measurements, also holds for the minimum. So
$$
\mathcal{D}(A,B) = \min_{\{\Pi_j\}}\tilde S_{\{\Pi_j\}}(A|B) - S(A|B) \geq 0.
$$

This theorem shows that quantum discord and SSA are intimately connected, in that the existence of the former is guaranteed by the validity of the latter, and the nullity of the former is guaranteed by the saturation of the latter. We have shown that for any bipartite state, we can introduce a third system that executes a measurement on one of the subsystems, and the resulting tripartite system allows us to investigate the role of quantum discord in quantum information theory.

\section{Interpreting quantum discord through quantum state merging  via SSA}
Quantum state merging protocol is the extension of the classical Slepian-Wolf protocol~\cite{thomascover} into the quantum domain where Alice and Bob share the quantum state $\rho_{AB}^{\otimes n}$, with each party having the marginal density operators $\rho_{A}^{\otimes n}$ and $\rho_{B}^{\otimes n}$ respectively. Let $\ket{\Psi_{ABC}}$ be a purification of $\rho_{AB}.$ Assume, without loss of generality, that Bob holds $C.$ The quantum state merging protocol quantifies the minimum amount of quantum information which Alice must send to Bob so that he ends up with a state arbitrarily close to $\ket{\Psi}_{B'BC}^{\otimes n},$ $B'$ being a register at Bob's end to store the qubits received from Alice. It was shown that in the limit of $n\rightarrow \infty$, and asymptotically vanishing errors, the answer is given by the quantum conditional entropy~\cite{how07}: $S(A|B) = S(A,B) - S(B)$. When $S(A|B)$ is negative, Bob obtains the full state with just local operations and classical communication, and distill $-S(A|B)$ ebits with Alice, which can be used to transfer additional quantum information in the future.

That quantum discord has an operational interpretation in terms of quantum state merging was shown in~\cite{md10,cabmpw10}. In ~\cite{md10}, it was shown that quantum discord is the markup in the cost of quantum communication in the process of quantum state merging, if one discards relevant prior information. SSA served as a crucial link in this exercise. An intuitive argument for the above interpretation of quantum discord can be made through strong subadditivity, which can also be written as~\cite{how07}
\be
\label{eq:ssa}
S(A|B,C) \leq S(A|B).
\ee
From the point of view of the state merging protocol, the above has a very clear interpretation: having more prior information makes state merging cheaper. Or in other words, throwing away information will make state merging more expensive. Thus, if Bob discards system C, it will increase the cost of quantum communication needed by Alice in order to merge her state with Bob. Our goal now is to take these results a step further. In particular, we show that we can interpret quantum discord in terms of the mother protocol~\cite{adhw09} and thus elucidate its connection to all the children protocols that can be derived from the mother.

\section{The mother protocol} % (fold)
%\label{sec:quantum_metrology_with_obbo_states}
We begin by briefly describing the Mother protocol and its generalization to the fully quantum Slepian Wolf (FQSW) protocol. The mother protocol~\cite{adhw09} is a transformation of a quantum state $(\ket{\Psi^{ABR}})^{\otimes n}$. At the start, Alice holds the $A$ shares and Bob the $B$ shares. The reference system $R$ is purifying the $AB$ system and does not actively participate in the protocol. The Mother protocol can be viewed as an entanglement distillation between $A$ and $B$ when the only type of communication permitted is the ability to send qubits from Alice to Bob. The transformation can be expressed concisely in the resource inequality formalism as~\cite{dhw08}

\begin{equation}
\langle\Psi^{AB}\rangle  + \frac{1}{2} I (A:R) [q \rightarrow q] \geq \frac{1}{2} I (A:B)[qq].
\end{equation}
The above inequality means that $n$ copies of the state $\Psi$ can be converted to $\frac{1}{2} I (A:B)$ EPR pairs ($[qq]$) per copy, provided Alice is allowed to communicate with Bob by sending him qubits at the rate $\frac{1}{2} I (A:R)$ ($[q\rightarrow q]$) per copy.

One can generalize the mother protocol to a stronger inequality known as the FQSW protocol. This inequality states that starting from the state  $(\ket{\Psi^{ABR}})^{\otimes n}$, and using $\frac{1}{2} I (A:R)$ bits of quantum communication from Alice to Bob, they can distill
$\frac{1}{2} I (A:B)$ EPR pairs per copy, and in addition Alice can accomplish merging her state with Bob. In the process of accomplishing state merging, they create the state $(\ket{\Phi^{\hat{B}R}})^{\otimes n}$, where $\hat{B}$ is a register held with $B$ and  $\Psi^{R}$ $=$  $\Phi^{R}$. Since all purifications are equivalent up to local unitaries, Bob can convert  $\Phi^{\hat{B}}$ to $\Psi^{AB}$ at his end and thus complete the state merging with Alice.  Hence in the state merging task, as described above, Alice is able to successfully transfer her entanglement with the reference system $R$ to Bob. Writing the FQSW in terms of a resource inequality

\be
\langle \mathcal{W}^{S \rightarrow AB} : \Psi^{S}\rangle  + \frac{1}{2} I (A:R) [q \rightarrow q] \geq \frac{1}{2} I (A:B)[qq] + \langle id^{S \rightarrow \hat{B}} : \Psi^{S}\rangle.
\ee
The above inequality is another way of expressing the FQSW protocol, where we accomplish state merging as well as entanglement distillation.
The state $S$ on the left-hand side of the inequality, is distributes to Alice and Bob, while on the right-hand side, that same state is given to Bob alone. $\mathcal{W}^{S \rightarrow AB}$ is an isometry taking the system $S$ to $AB$~\cite{adhw09}. The FQSW protocol is valid asymptotically in the limit of a large number of copies and this is denoted by the symbol $\geq$.

\subsection{Quantum state merging primitive from FQSW}

We start by expressing the quantum state merging protocol~\cite{how07} as a resource inequality
\begin{equation}
\langle\Psi^{AB}\rangle  + S(A|B) [q \rightarrow q]  + I (A:B)_{\psi} [c \rightarrow c] \geq \langle id^{S \rightarrow \hat{B}} : \Psi^{S}\rangle.
\end{equation}
This accomplishes state merging from Alice to Bob at the cost of $S( A|B)$ bits of quantum communication. In the case when $S(A|B)$ is negative, Alice and Bob can distill this amount of entanglement in the form of Bell pairs. Quantum state merging this provides an operational interpretation of $S(A|B)$, assuming we ignore the amount of classical communication needed to accomplish the required state merging.

We can derive quantum state merging from the FQSW if the entanglement produced at the end of the FQSW protocol, can be used to perform teleportation. We can see this through simple manipulation of the resource inequalities described above. We start by describing quantum teleportation as
\be
 [qq]  + 2 [c \rightarrow c] \succeq  [q \rightarrow q].
\ee
It means that one requires a shared ebit and two bits of classical communication to accomplish teleportation of a quantum state. The symbol $\succeq $ is used to denote exact attainability as compared to $\geq$ which is to denote asymptotic attainability. From the FQSW protocol (eq 2.3), using the entanglement produced at the end for quantum communication (eq 2.5), one gets the quantum state merging primitive (eq 2.4).

\section{Quantum discord as a measure of the coherence of the Mother protocol}
In this section we present our main result, that quantum discord is a measure of how coherently the mother protocol is performed between two parties, Alice and Bob. More specifically, we will study the loss of information and coherence at Bob's end. To that end, we consider arbitrary quantum operations to model decoherence. We also consider a quantum operation where quantum measurements are made at Bob's end and the results are discarded. In practice, such a pre-measurement state can be due to the environment assisted decoherence.

To begin, expand the size of the Hilbert space so that an arbitrary pre-measurement (or any other quantum operation) can be modeled by coupling to the auxiliary subsystem and then discarding it. We assume $C$ to initially be in a pure state $\ket{\mathbf{0}}$, and a unitary interaction $U$ between $B$ and $C$. Letting primes denote the state of the system after $U$ has acted, we have $S(A,B) = S(A,BC)$ as $C$ starts out in a product state with $AB$. We also have $I(A : BC) = I(A' : B'C')$. As discarding quantum systems cannot increase the mutual information, we get $I(A' : B') \leq I(A' : B'C')$. Now consider the FQSW protocol between A and B in the presence of $C$. We have $S(A|B) = S(A) - I(A : B) = S(A) - I(A : BC) = S(A|BC).$ After the application of the unitary $U$, but before discarding the subsystem $C$, the cost of merging is still given by $S(A'|B'C') = S(A|B)$. This implies that one can always view the cost of merging the state of system $A$ with $B$,  as the cost of merging $A$ with the system $BC$, where $C$ is some ancilla (initially in a pure state) with which $B$ interacts coherently through a unitary $U$. Such a scheme does not change the cost of state merging, as shown, but helps us in counting resources. Discarding system $C$ yields
\be
\label{eq:Idiff}
I(A' : B') \leq I(A':B'C') = I(A:BC) = I(A : B),
\ee
or alternatively,
\be
\label{eq:Sdiff}
S(A'|B') \geq S(A'|B'C') = S(A|B).
\ee
Now consider a protocol which we call as $FQSWD_{B},$ (Fully Quantum Slepian Wolf after decoherence) where the subscript refers to the decoherence at B. The resource inequality for $FQSWD_{B}$ is
\begin{equation}
\label{eq:fqswdb}
\langle \mathcal{U}^{S \rightarrow A'B'} : \Psi^{S}\rangle  + \frac{1}{2} I (A':R') [q \rightarrow q] \geq \frac{1}{2} I (A':B')[qq] + \langle id^{S \rightarrow \hat{B}} : \Psi^{S}\rangle.
\end{equation}
As in the fully coherent version, the Alice is able to transfer her entanglement with the reference system $R$', and is able to distill $\frac{1}{2} I (A':B')[qq]$ EPR
pairs with Bob. Since $I(A' : R') = I(A : R),$ the FQSW protocol without decoherence has a net gain of $\frac{1}{2} I (A:B)[qq] -\frac{1}{2} I (A':B')[qq] = \frac{1}{2}D$
EPR pairs. Since the cost of quantum communication is same for both ( coherent and decohered) versions, we regard $D$ as the metric of how coherently the FQSW protocol takes place. We now show that the minimum of $D$ over all possible measurements is the quantum discord $\mathcal{D}$. The state $\rho_{AB},$ under measurement of subsystem $B$, changes to $\rho_{AB} ' = \sum_{j} p_{j} \rho_{A|j} \otimes \pi_j,$ where $ \{\pi_j\} $ are orthogonal projectors resulting from a Neumark extension of the POVM elements. The unconditioned post measurement states of $A$ and $B$ are
$$
\rho_{A}' =\sum_{j} p_{j} \rho_{A|j} = \rho_{A},~~~\rho_{B}' = \sum_{j} p_{j} \pi_j.
$$ Computing the value of $I(A':B')$, we get
\ben
I(A':B') &=&  S(A') +S(B') - S(A',B'),\nonumber  \\
        &=& S(A') + H(p) -\big\{H(p) + \sum_{j} p_{j} S(\rho_{A|j})\big\}, \nonumber\\
            &=& S(A) - \sum_{j} p_{j} S(\rho_{A|j}).
\een
After maximization, it reduces to $\mathcal{J}(\rho_{AB})$, as defined earlier as is the reduction to rank 1 POVMs. One might consider reverting to Zurek's original definition of quantum discord, which incidentally first appeared in~\cite{z00}. Then one does not have to throw in the maximization; quantum discord quantifies the increase in quantum state merging due to environmental projection, and hence the quantity $\mathcal{D}$ serves as a valid measure for the net loss in the number of EPR pairs in the mother protocol due to the environment. The above connection between quantum discord and the mother protocol suggests that quantum discord can also serve as a measure of coherence in accomplishing any of the children protocols that can be derived from the mother. We illustrate this in the next section.

\section{Role of quantum discord in the children protocols}
In this section we see that by connecting quantum discord with the FQSW protocol, we can interpret quantum discord as the advantage of quantum coherence in various scenarios. There in lies the power of our approach.

\subsection{Quantum discord as the mark up in the cost of quantum communication to state merging}

We can easily derive the results of ~\cite{md10} from the previous section. In particular, consider the $QSMD_{B}$ (Quantum state merging protocol with decoherence at party $B$). One can get $QSMD_{B}$ from $FQSWD_{B}$ if one recycles the entanglement produced at the end of the $FQSWD_{B}$ protocol to perform quantum teleportation. We start by expressing $QSMD_{B}$ in the form of a resource inequality,

\begin{equation}
\langle\Psi^{A'B'}\rangle  + S(A'|B') [q \rightarrow q]  + I (A':B')_{\psi} [c \rightarrow c] \geq \langle id^{S \rightarrow \hat{B}} : \Psi^{S}\rangle.
\end{equation}

The cost of quantum communication in this case is $S(A'|B')$. Thus the mark up in this cost is $S(A'|B') - S(A|B)$, which is equal to the quantum discord of the original state.

\subsection{Quantum discord and noisy super-dense coding}
The noisy super-dense coding can be derived by combining the mother with super-dense coding~\cite{superdense}. It can be expressed in the form of the resource inequality as,

\begin{equation}
\langle\Psi^{AB}\rangle  + S(A) [q \rightarrow q] \geq I (A:B)[c \rightarrow c].
\end{equation}
When the party $B$ is undergoing decoherence, the noisy superdense coding can be expressed as,

\begin{equation}
\langle\Psi^{A'B'}\rangle  + S(A') [q \rightarrow q] \geq I (A':B')[c \rightarrow c].
\end{equation}
We note that $ S(A)  = S(A') $.
Thus, due to decoherence, the number of classical bits communicated through this protocol gets reduced by the amount $I (A:B) -  I (A':B')$, which is equal to the quantum discord of the original state.

\subsection{Quantum discord and entanglement distillation}

The one way entanglement distillation can be expressed as~\cite{distill},
\begin{equation}
\langle\Psi^{AB}\rangle  + I (A:R) [c \rightarrow c] \geq I (A\rangle B)[qq].
\end{equation}
In the above equation, $I(A\rangle B)[qq] = -S(A|B)$, and is also known as the coherent information~\cite{cohinfo}. When the party $B$ is undergoing decoherence we get,

\begin{equation}
\langle\Psi^{A'B'}\rangle  + I (A':R') [c \rightarrow c] \geq I (A'\rangle B')[qq].
\end{equation}
Since, $ I (A':R') = I(A:R)$, we see that the net loss in entanglement distillation is equal to $S(A'|B') - S(A|B)$ which again is the quantum discord of the original state.

\section{Conclusion} % (fold)

Our work elucidates the role non classical correlations, those captured by quantum discord, play in quantum information processing tasks. For an important class of problems described above, quantum discord is shown to be a measure of how coherently the protocol was performed. We have quantified the cost due to decoherence we suffer in quantum communication protocols and this is aptly captured by quantum discord. We hope that this work places quantum discord at the heart of quantum information theory, and demonstrates the vital role it plays in quantifying the cost of decoherence in almost all quantum information processing protocols.

\section*{Acknowledgements}
This work was supported in part by the  EPSRC (EP/H03031X/1), the EC integrated project Q-ESSENCE, US European Office of Aerospace Research (FA8655-09-1-3020), and the Center for Quantum Information and Control (CQuIC) where this work was done, and NSF Grant Nos. 0903953 and 0903692.

\end{document}